\documentclass{elsart4-1}

\usepackage{amssymb}
\usepackage[latin1]{inputenc}
\usepackage[english,french]{babel}
\usepackage{graphicx,psfrag}

\newtheorem{e-proposition}[theorem]{Proposition}

\newtheorem{e-definition}[theorem]{Definition\rm}

\renewcommand{\theequation}{\arabic{equation}}
\newcommand{\C}{{\mathrm c}}
\newcommand{\D}{{\mathrm d}}
\newcommand{\E}{\mathrm{e}}
\newcommand{\F}{\mathrm{f}}
\newcommand{\I}{\mathrm{i}}
\newcommand{\DD}{\mathcal{D}}
\newcommand{\FF}{\mathcal{F}}
\newcommand{\HH}{\mathcal{H}}
\newcommand{\LL}{\mathcal{L}}
\newcommand{\NN}{\mathcal{N}}
\newcommand{\PP}{\mathcal{P}}
\newcommand{\Sc}{\mathcal{S}}

\newcommand{\tf}{{t_\mathrm{f}}}
\newcommand{\average}[1]{\left<{#1}\right>}

\newcommand{\kb}{k_\mathrm{B}}
\newcommand{\derpart}[2]{\frac{\partial #1}{\partial #2}}
\newcommand{\pq}[1]{\left[{#1}\right]}

\def\og{\leavevmode\raise.3ex\hbox{$\scriptscriptstyle\langle\!\langle$~}}
\def\fg{\leavevmode\raise.3ex\hbox{~$\!\scriptscriptstyle\,\rangle\!\rangle$}}

\begin{document}

\centerline{Physics}
\begin{frontmatter}


\title{Work and heat probability distributions in out-of-equilibrium systems}
 \author{Alberto Imparato}
 \ead{alberto.imparato@polito.it}
\address{Dipartimento di
Fisica, INFN-Sezione di Torino, CNISM-Sezione di Torino, Politecnico
di Torino, C.so Duca degli Abruzzi 24, 10129 Torino, Italy}
 \author{Luca Peliti}
 \ead{peliti@na.infn.it}
\address{Dipartimento di Scienze
Fisiche, INFN-Sezione di Napoli, CNISM-Sezione di Napoli, Università
``Federico~II'', Complesso Monte S. Angelo, 80126 Napoli, Italy}
\selectlanguage{english}

\medskip
\begin{center}
{\small November 24, 2006}
\end{center}

\begin{abstract}
We review and discuss the equations governing the distribution of
work done on a system which is driven out of equilibrium by external
manipulation, as well as those governing the entropy flow to a
reservoir in a nonequilibrium system. We take advantage of these
equations to investigate the path phase transition in a manipulated
mean-field Ising model and the large-deviation function
for the heat flow in the asymmetric exclusion process with
periodically varying transition probabilities.
{\it To cite this article: A. Imparato, L. Peliti, C. R.
Physique 6 (2005).}

\vskip 0.5\baselineskip

\selectlanguage{francais}
\noindent{\bf R\'esum\'e}
\vskip 0.5\baselineskip
\noindent
{\bf Distributions du travail et de la chaleur dans des systèmes hors
équilibre.}
Nous passons en revue et discutons les équations régissant
la distribution du travail effectué sur un système
manipulé hors d'équilibre, ainsi que celles qui régissent
le flux d'entropie vers un reservoir dans un système hors
d'équilibre. Nous exploitons ces équations dans l'étude
de la transition de phase dans les chemins d'un modèle d'Ising
champ moyen manipulé et de la fonction des grandes déviations
pour le flux d'entropie dans le modèle d'exclusion asymétrique à
probabilités de transition périodiques dans le temps.
{\it Pour citer cet article~: A. Imparato, L. Peliti, C. R.
Physique 6 (2005).}

 \vskip 0.5\baselineskip
\noindent{\small{\it Mots-cl\'es~:} Procès hors équilibre~; Distribution du travail~; Flux d'entropie}}
\end{abstract}
\end{frontmatter}


\selectlanguage{english}
\section{Introduction}
In the recent years we have seen an important outburst of
activity in the field of nonequilibrium thermodynamics and
statistical mechanics, sparked by the discovery of a number of
results of remarkable generality and impact~\cite{Bochkov1981a,Bochkov1981b,EvansCohen93,%
EvansSearles95,GallCoh95,Jarzynski97,Kurchan98,LebSpohn99,%
Crooks99,Crooks00,Hummer01,hatano-2001-86,Maes03,Gaspard04,Gaspard05,Seifert05,Imparato06,cleuren-2006-96}.
Some of these results yield predictions on the properties
of the distribution of the work performed on a system as it is
manipulated, while others describe properties of the
distribution of the entropy generated in the nonequilibrium process.
It is clear that in order to make good use of this
information, it is advisable to investigate in detail the
properties of the distribution of both work and heat
in a manipulated system.

In the present Contribution we discuss the work and the heat distribution
in some systems, pointing out some aspects that we have
found interesting enough to be brought to the attention of our colleagues.
We first consider the expression of the generating function for
the work distribution in a ``large'' system. Although our results formally
hold in the limit of infinite size, they are of
interest also for systems small enough that the relevant energy barriers are of
the order of a few $\kb T$, well within the range in which
work fluctuations can be observed. In a large system, as already pointed out
by Ritort~\cite{Ritort2state04} and discussed in~\cite{Imparato05}, the generating function of the work
distribution is dominated by the contribution of phase space paths
which satisfy an ordinary differential equation akin to a classical
equation of motion. We have found that in some situations these paths
can exhibit a singularity, for some protocols, as a function of
the variable conjugate to the accumulated work. This singularity is similar
to a phase transition taking place in the space of paths. We shall exhibit
a simple model in which this phenomenon takes place and attempt to define
the corresponding phase diagram.

We also consider the heat flow distribution in a general Markov process,
whose differential equations were derived by Lebowitz
and~Spohn~\cite{LebSpohn99}. We were able to evaluate the
solution of this equation for the asymmetric exclusion model
with periodically varying transition probabilities, and to exhibit
that the Gallavotti-Cohen~\cite{GallCoh95} symmetry also holds in this case.

\section{The generating function for the work distribution}
We first briefly review the derivation of the generating
function for the work distribution of manipulated systems,
which may be found in~\cite{Imparato05}.
Let us consider a system with discrete states $i$ of
energy $H_i(\mu)$, where $\mu$ is a parameter which is
manipulated according to some protocol $\mu(t)$, starting at
$t=0$. The evolution of the system is described by
a markovian stochastic process: given, for any two states $i$, $j$, the
transition rate $k_{ij}(t)$ from state $j$ to state $i$ at time
$t$, the system satisfies the set of differential equations
\begin{equation}
    \derpart{p_i}{t}= \sum_{j(\ne i)} \left[k_{ij}(t) p_j(t)
    - k_{ji}(t) p_i(t)\right], \label{evp}
\end{equation}
where $ p_i(t)$ is the probability that the system is found at
state $i$ at time $t$. Let $p^\mathrm{eq}_i(\mu)$ represent the
canonical equilibrium distribution corresponding to a given value of $\mu$.
We have
\begin{equation}
    p^\mathrm{eq}_i(\mu)=\frac{\E^{-\beta H_i(\mu)}}{Z_\mu},
\end{equation}
where $Z_\mu=\sum_i \E^{-\beta H_i(\mu)}=\E^{-\beta F_\mu}$ is the partition
function corresponding to the value $\mu$ of the parameter, and $F_\mu$
the relative free energy. The transition rates $k_{ij}(t)$
are compatible with
the equilibrium distribution $p^\mathrm{eq}_i(\mu)$, i.e., one has, for all
$i$,
\begin{equation}
    \sum_{j (\neq i)}\left[k_{ij} (t) p^{\mathrm{eq}}_j(\mu(t))
    - k_{ji} (t) p^{\mathrm{eq}}_i(\mu(t))\right]
   =0.
\label{equil}
\end{equation}
We assume that the system is at equilibrium at $t=0$, and
therefore, that $p_i(t)$ satisfies the initial condition
\begin{equation}
    p_i(t{=}0)=p^\mathrm{eq}_i(\mu(0)).
\end{equation}
We now consider
the joint probability distribution $\Phi_i(W,t)$ that the system
is found in state $i$, having received a work $W$, at time $t$. The
set of differential equations satisfied by the distribution
functions $\Phi_i(W,t)$ reads
\begin{equation}
    \derpart{\Phi_i}{t}=\sum_{j(\ne i)}\left[ k_{ij} (t)
    \Phi_j(W,t) - k_{ji} (t) \Phi_i(W,t)\right]-\dot \mu H'_i(\mu(t))\,\derpart{  \Phi_i}{W}. \label{eqphi}
\end{equation}
The joint probability distribution $\Phi_i(W,t)$ satisfies the
initial condition
\begin{math}
    \Phi_i(W,0)=\delta(W)\,p^\mathrm{eq}_i(\mu(0)).
\end{math}
Then the state-independent
work probability distribution $P(W,t)$ is defined by
\begin{math}
    P(W,t)=\sum_i\Phi_i(W,t).
\end{math}
It is convenient to introduce the generating function of $\Phi_i$
with respect to the work distribution, defined by
\begin{equation}
    \Psi_i(\lambda,t)=\int \D W \;\E^{\lambda W} \Phi_i(W,t).
\label{defpsi}
\end{equation}
We assume that $\Phi_i(W,t)$ vanishes fast enough, as
$|W|\to\infty$, for $\Psi_i(\lambda,t)$ to exist for any
$\lambda$. The function $\Psi_i$ satisfies, for all $\lambda$,
the initial condition
\begin{math}
    \Psi_i(\lambda,t_0)=p^{\mathrm{eq}}_i(\mu(0)),
\end{math}
and evolves according to the differential equation
\begin{equation}
   \partial_t \Psi_i(\lambda,t)=
\sum_{j(\ne i)} \left[k_{ij} \Psi_j -k_{ji} \Psi_i\right]
     +\lambda \dot
    \mu\,\frac{\partial H_i(\mu(t))}{\partial \mu}\,
    \Psi_i(\lambda, t). \label{detpsi}
\end{equation}
Exploiting (\ref{equil}), it is easy
to verify that, if $\lambda=-\beta$,
the solution of (\ref{detpsi}) with its initial condition
reads
\begin{equation}\label{psimt}
    \Psi_i(-\beta, t)=\frac{\E^{-\beta H_i(\mu(t))}}{Z_{\mu(0)}}
    =\frac{Z_{\mu(t)}}{Z_{\mu(0)}}\,p^\mathrm{eq}_i(\mu(t)).
\end{equation}
We can thus straightforwardly check that the
solution of (\ref{detpsi}) verifies the Jarzynski equality~\cite{Jarzynski97}:
\begin{equation}
    \average{\E^{-\beta W}}=\frac{Z_{\mu(t)}}{Z_{\mu(0)}}
    =\E^{-\beta\left(F(\mu(t))-F(\mu(0))\right)}.\label{jarzder}
\end{equation}
It is thus possible, in principle, to evaluate the probability
distribution function of the work $W$ by solving the equations
(\ref{eqphi}) or (\ref{detpsi}) for all the microscopic states
$i$. This approach has been implemented in \cite{Imparato05a} for a
simple model of a biopolymer.

The approach discussed in the previous section becomes quickly
unwieldy as the complexity of the system increases: the dimension
of the system (\ref{eqphi}) is equal to the number of microscopic
states of the system.  When the
system considered is characterized by a large number of degrees of
freedom, it is convenient to introduce some collective variables, and
an effective free energy, in order to reduce the complexity of the
problem. The assumption underlying this approach is that the
system reaches on a comparatively short time scale a
quasiequilibrium state constrained by the instantaneous value of
the collective coordinate. Thus, on the the time scale of the
experiment, the state of the system can be well summarized by the
collective coordinate, with the corresponding free energy playing
the role of the hamiltonian.

Let us now consider a system
characterized by a generic equilibrium free energy function
$\FF_\mu(M)$, where $\mu$ is again the parameter which is
manipulated, and $M$ is some collective (mean-field) variable.
We assume that the system dynamics is stochastic and
markovian: let $P(M,t)$ denote the probability distribution
function of the variable $M$ at time $t$, then its time evolution
is described by the differential equation
\begin{equation}
    \derpart{P}{t}=\widehat\HH\,P,
\end{equation}
where $\widehat\HH$ is a differential operator which depends
on the parameter $\mu$. We require that the operator
$\widehat\HH$ is compatible with the equilibrium distribution
function of the system, i.e., that the relation
\begin{equation}\label{equilibrium}
    \widehat\HH\, \E^{-\beta \FF_\mu(M)}=0
\end{equation}
holds for any value of $\mu$.

The work done on a system during the manipulation, along a given
stochastic trajectory $M(t)$, is given by
\begin{equation}
    W=\int_0^{t} \D t'\,  \dot \mu(t') \,
    \frac{\partial \FF_\mu(M(t'))}{\partial \mu}\, .
\end{equation}
Using the same arguments as for the discrete case, one finds that
the time evolution of the joint probability distribution
$\Phi(M,W,t)$ of $M$ and $W$ is described by the differential
equation
\begin{equation}
    \label{phi:eq}
    \frac{\partial \Phi}{\partial t}=\widehat \HH \Phi
    -\dot \mu \frac{\partial \FF_\mu}{\partial \mu}
    \frac{\partial\Phi}{\partial W}.
\end{equation}
It can be easily shown that the solution of (\ref{phi:eq})
 identically satisfies the Jarzynski equality~\cite{Imparato05b}.

Equation (\ref{phi:eq}) becomes much easier to treat if one
introduces the generating function $\Psi(M,\lambda,t)$ for the
work distribution:
\begin{equation}
    \label{psi:def}
    \Psi(M,\lambda,t)=\int \D W\,\E^{\lambda W}\,\Phi(M,W,t).
\end{equation}
Equation (\ref{phi:eq})  thus becomes
\begin{equation}
    \label{psi:eq}
    \frac{\partial \Psi}{\partial t}=\widehat \HH \Psi
    +\lambda\dot \mu \frac{\partial \FF_\mu}{\partial \mu}\Psi,
\end{equation}
with the initial condition
\begin{math}
    \Psi(M,\lambda, 0)=p^{\mathrm{eq}}_{\mu(0)}(M)=\E^{-\beta \FF_{\mu(0)}(M)}/{Z_{\mu(0)}}.
\end{math}

As shown in~\cite{Imparato05b}, one can derive a path integral representation of the solution of
(\ref{psi:eq}), taking for the differential operator
$\widehat\HH$ the expression
\begin{equation}
\label{diffop:def}
    \widehat \HH\cdot{}=\sum_{k=0}^\infty \frac{\partial^k}{\partial M^k}
    \left\{g_k(M)\cdot{}\right\}.
\end{equation}
(The coefficients $g_k(M)$ also depend on $\mu$, but this
dependence is understood to lighten the notation.) One obtains the formal solution
\begin{equation}
\label{integral:def}
    \Psi(M,\lambda,\tf)=\int\D M_0
    \int_{M(0)=M_0}^{M(\tf)=M} \DD\gamma\DD M\;\exp\left\{\Sc[\gamma,M]\right\}\;\Psi(M_0,\lambda,0),
\label{path}
\end{equation}
where
\begin{math}
    \Sc[\gamma,M]=\int_0^\tf\D t\;\LL(t).
\end{math}
The ``lagrangian'' $\LL$ is given by
\begin{equation}
    \LL(t)=\left.\left(\gamma \dot M+\HH(\gamma,M)
    +\lambda \dot \mu \,
    \frac{\partial \FF_\mu}{\partial \mu}\right)
    \right|_{\gamma(t),M(t),\mu(t)},
\label{LL}
\end{equation}
where the ``hamiltonian'' $\HH(\gamma,M)$ is defined by
\begin{equation}
    \HH(\gamma,M)=\sum_{k=0}^\infty \gamma^k g_k(M).
\end{equation}
Let $N$ indicate the size of the system, and let us define the
``intensive quantity'' $m=M/N$. We can thus define, in the
thermodynamic limit $N\to\infty$, $m=\mbox{const.}$, the densities
\begin{equation}
    f_\mu(m)=\lim_{N\to\infty}\frac{\FF_\mu(N m)}{N},\qquad
    H(\gamma,m)=\lim_{N\to\infty}\frac{\HH(\gamma, Nm)}{N},\qquad
    \ell(t)=\lim_{N\to\infty}\frac{\LL(t)}{N},\label{smallH}
\end{equation}
with $\ell(t)=\gamma \dot m+H(\gamma,m)
    +\lambda \dot \mu \,
    {\partial f_\mu}/{\partial \mu}.$
As discussed in \cite{Ritort2state04,Imparato05b}, when the system size $N$ is large
enough, the path integral in (\ref{path}) is dominated by the
classical path $(\gamma_\C(t),m_\C(t))$, solution of the
equations
\begin{equation}
    \frac{\delta \Sc}{\delta \gamma(t)}=0 \Longrightarrow \dot m
    =-\frac{\partial H}{\partial \gamma};\qquad
    \frac{\delta \Sc}{\delta m(t)}=0\Longrightarrow \dot
    \gamma=\frac{\partial H}{\partial m}+\lambda \dot \mu
    \frac{\partial^2 f_\mu}{\partial m\partial \mu}.\label{eq2}
\end{equation}
We shall now see that by requiring the system is in
equilibrium before the manipulation starts, we impose an initial
condition on these equations. In order to evaluate the integral
over $M_0$ in (\ref{path}) with the saddle-point method, we
note that $\Psi(M,\lambda,0)$ appearing on its rhs, is
given by \begin{math} p^{\mathrm{eq}}_{\mu(0)}(M). \end{math}
Furthermore, from the definition of $\ell(t)$,
(\ref{smallH}), it follows that
\begin{equation}
    \int_0^\tf \D t\,  \ell(t)=m_\tf \gamma_\tf-m_0 \gamma_0
    +\int_0^\tf \D t\, \left[-\dot \gamma m+ H +\lambda \dot \mu
\partial_\mu f_\mu\right].
\label{cambv}
\end{equation}
Thus, substituting (\ref{cambv}) into (\ref{path}), and
taking the derivative with respect to $m_0=M_0/N$,  we obtain the
saddle-point condition
\begin{equation}
\gamma(t{=}0)=-\beta \left.\frac{\partial f_\mu}{\partial m}\right|_{t=0}.
\label{gamma0}
\end{equation}

In this way one can devise a strategy to evaluate
$\Psi(M,\lambda,\tf)$ for a given manipulation protocol $\mu(t)$,
when the system size $N$ is large enough. One has to solve the
classical evolution equations (\ref{eq2}) with a
two-point boundary condition: namely, (\ref{gamma0}) should be
imposed at $t=0$, and the condition $Nm(\tf)=M$ should be imposed at
the final time $\tf$. Once the relevant classical path
$(\gamma_\C(t),m_\C(t))$ has been evaluated, one can obtain
the action density
$s[\gamma_\C,m_\C]=\lim_{N\to\infty}\Sc[\gamma_\C,N
m_\C]/N$ from the expression
\begin{math}
    s[\gamma_\C,m_\C]=\int_0^\tf\D t\;\ell(t).
\end{math}
We are interested in the state-independent
work probability distribution
\begin{equation}
    P(W,\tf)=\int \D \lambda\,  \E^{-\lambda W} \,\Gamma(\lambda,\tf),
    \label{pw}
\end{equation}
where we have defined
\begin{math}
    \Gamma(\lambda,\tf)=\int \D M\; \Psi(M,\lambda,\tf).
\end{math}
We shall now see that evaluating $\Gamma(\lambda,\tf)$ identifies a
well-defined boundary condition on $\gamma_\C (\tf)$. We have
indeed
\begin{equation}
    \Gamma(\lambda,\tf) = \int \D M\, \D M_0 \int_{M(0)=M_0}^{M(\tf)=M}
  \DD\gamma\DD M\;\exp\left[N \int \D t\, \ell(t)\right]
    \Psi(M_0,\lambda,0). \label{defg}
\end{equation}
In order to evaluate the integral over $M$ with the saddle point
method, we notice that, upon derivation of the rhs of
(\ref{cambv}) with respect to $m_\tf$, we obtain the condition
\begin{math}
    \gamma_\F\equiv\gamma(\tf)=0.
\end{math}
Thus, the equation of motions (\ref{eq2})  have to
be solved with the initial and the final conditions that we have derived:
let $(\gamma_\C^*(t),m_\C^*(t))$ denote
the solution of equations~(\ref{eq2}) satisfying these
conditions. For each value of $\lambda$, taking into account its
initial condition, we obtain the following saddle point
estimation for $\Gamma(\lambda,\tf)$:
\begin{equation}
    \Gamma(\lambda,\tf)\propto \frac{\exp\left[N g(\lambda)\right]} {Z_0},
\label{Gamma_g}
\end{equation}
where
\begin{equation}
    g(\lambda)=\int_0^\tf \D t\;\ell^*_\C(t)-\beta f_{\mu_0}(m_0^*).
    \label{defgibbs}
\end{equation}
In this equation, $\ell^*_\C(t)$ is $\ell(t)$ evaluated along the
classical path $(\gamma^*_\C(t),m^*_\C(t))$. In order to evaluate
the integral on the rhs of (\ref{pw}), we use the saddle point
method again, and obtain
\begin{equation}
    P(Nw,\tf)=\NN\exp\left\{N \left[-\lambda^*(w) w +g(\lambda^*(w))\right]\right\},
 \quad\mbox{ with }\quad   g'(\lambda^*)=w,\label{eqw:eq}
\end{equation}
where $\mathcal N$ is a normalization constant. Notice that the
saddle point estimate for $P(W,\tf)$ obtained in this way, implies
that the distribution becomes more and more sharply peaked around
its maximum value as $N\rightarrow\infty$. This is compatible with
the expectation that the work fluctuations becomes relatively
smaller as the size of the system increases, and in the limit
$N\to\infty$, which can be thought as the limit of a macroscopic
system, no work fluctuations are observed, and the work done on
the system during the manipulation takes one single value,
corresponding to the most probable value of $P(W,\tf)$.

\section{A path phase transition}\label{thermo:sec}
In this section we consider an Ising model in  mean-field approximation,
which evolves according to a Fokker-Planck equation, whose
differential operator reads
\begin{equation}
\label{FP}
    \widehat \HH \cdot {}=\omega_0N\,\frac{\partial}{\partial M}\left[
    \left(\frac{\partial \FF}{\partial M}\right)\cdot{}
    +\beta^{-1}\frac{\partial}{\partial M}\cdot{}\right],
\end{equation}
leading to the hamiltonian
\begin{equation}
  H(\gamma,m)=\omega_0\left[\gamma \left(\frac{\partial f}{\partial m}\right)
    +\beta^{-1} \gamma^2\right],
\end{equation}
where the free energy density $f(m)=\FF(Nm)/N$ is given by
\begin{equation}\label{freeen:eq}
  f(m)=-\frac{J}{2}m^2-hm+\beta^{-1}\left[\left(\frac{1+m}{2}\right)
    \log\left(\frac{1+m}{2}\right)+\left(\frac{1-m}{2}\right)
    \log\left(\frac{1-m}{2}\right)\right].
\end{equation}

We take the magnetic field $h(t)$ as the external parameter that varies with the time and drives the system out of equilibrium.
The equations of motion (\ref{eq2}) thus become
\begin{equation}
    \dot m =-\derpart{H}{\gamma}=-\omega_0 \derpart{f}{m}-2 \kb T
        \omega_0\gamma;\qquad
    \dot \gamma =\derpart{H}{m}+\lambda \dot \mu \frac{\partial^2
    f}{\partial m\partial \mu}=\omega_0\frac{\partial^2f}{\partial
    m^2}\gamma-\lambda \dot h, \label{eq2_lang}
\end{equation}
The magnetic field $h(t)$ is taken to vary according to the linear protocol
\begin{equation}
\label{manh:eq}
h(t)=h_0+(h_1-h_0)\frac{t}{\tf}; \qquad 0\le t \le \tf.
\end{equation}

\begin{figure}[hb]
\psfrag{w}{\large $w$}
\psfrag{pw}[cb][cb][1.]{$P$}
\psfrag{pw1}[bb][bb][1.]{$\hat P$}
\psfrag{a}{$(a)$}
\psfrag{b}{$(b)$}
\begin{center}
 \includegraphics[width=7cm]{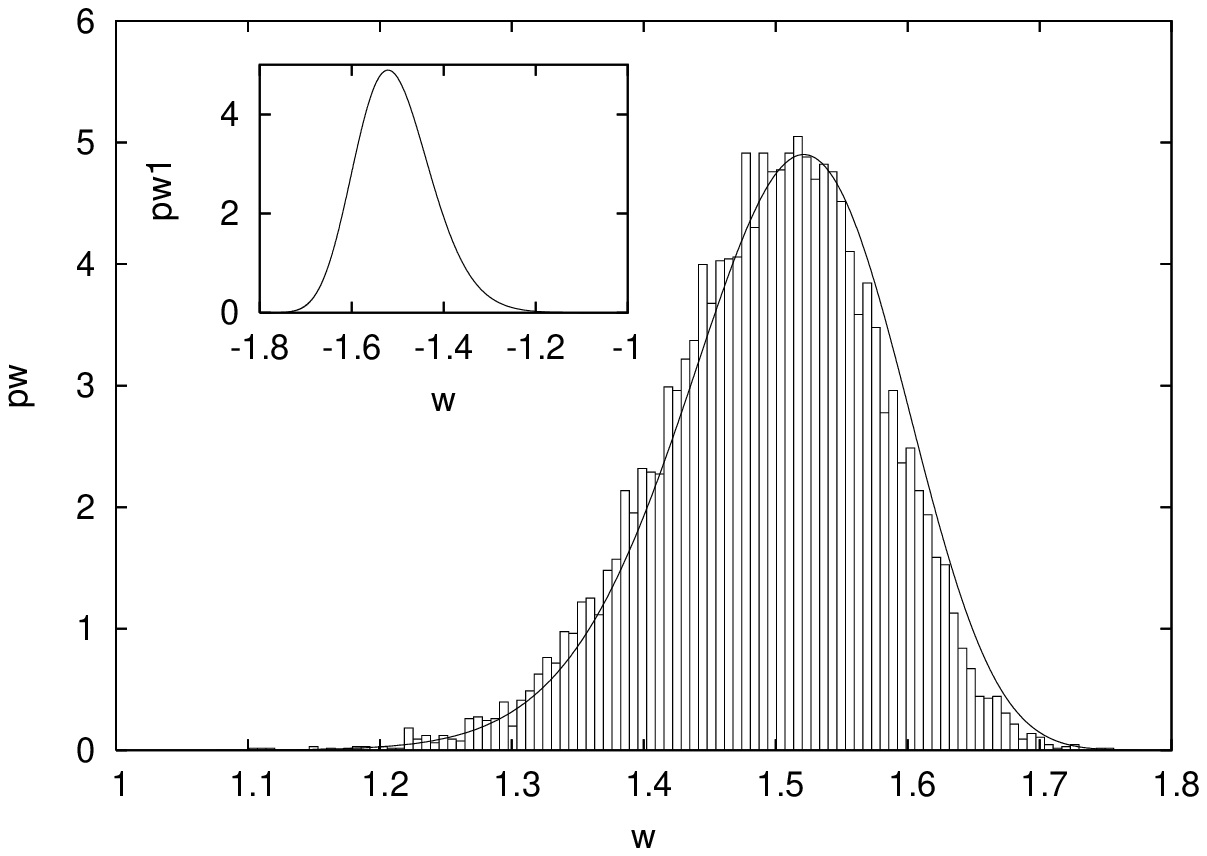}
 \includegraphics[width=7cm]{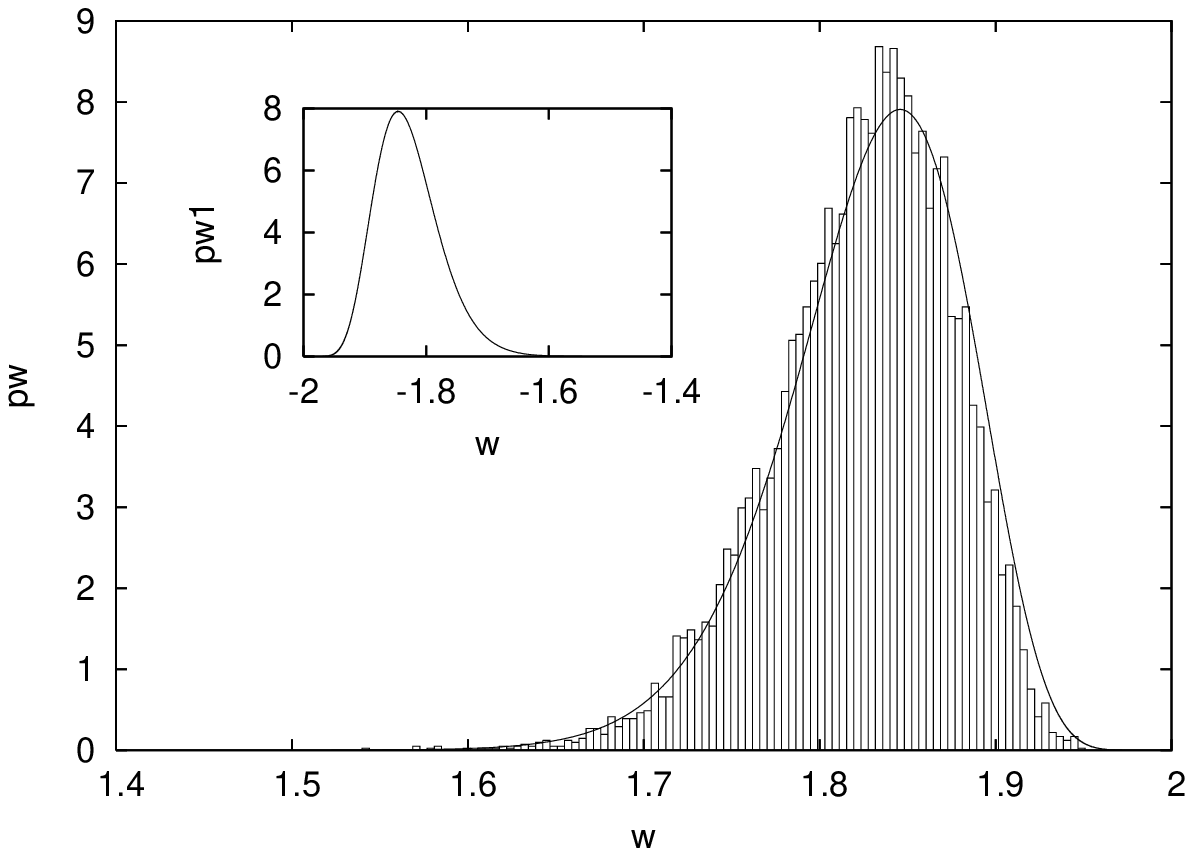}
\end{center}
 \caption{Probability distribution function $P(w)$ for the system described by
the differential operator (\ref{FP}) with equilibrium free energy
(\ref{freeen:eq}), manipulated according to the
protocol (\ref{manh:eq}), with $J_1=1.1$, $h_0=-h_1=-1$,
and $t_\mathrm{f}=2$ (left panel), $t_\mathrm{f}=4$ (right panel).  Continuous
line: probability density $P(w)$ of the work ``per spin''
$w=W/N$, with $N=100$.
The histogram of the work is obtained by 10000 simulations
of the process, see text. Insets:
 $\hat P(w)$ as given by (\ref{hatp}), whose
integral verifies the Jarzynski equality.}
\label{manh1}
 \end{figure}

Here we consider the case of Ising model below the transition temperature, i.e.,
$J=1.1$, and the initial and final value of the  magnetic field $h(t)$ are taken to be
$h_0=-h_1=-1$. In the present section we set $\beta=1/\kb T=1$.
We plot in figure~\ref{manh1} the probability distribution of the work done on the model, as obtained from~(\ref{eqw:eq}), for two values of the manipulation rate.
In the same figure the histograms of the work obtained by simulating the process are plotted. The process is simulated by  integrating the corresponding Langevin equation, using the Heun algorithm~\cite{Greiner88,Mannella02}. The agreement with the curves as obtained from~(\ref{eqw:eq}) is satisfactory.
In
the insets of the same figure, we plot the quantity $\hat P(w)$ defined by
\begin{equation}
\hat P(w)=\exp\pq{-\beta N w} P(w).
\label{hatp}
\end{equation}
On the
one hand we find $\int \D w \hat P(w)=\exp\pq{-\beta \Delta F}=1$
as predicted by the JE~(\ref{jarzder}), while on
the other hand the histogram obtained by the simulations exhibits
no point (no realization of the process) with $w<0=\Delta F$.
Thus the work
distribution obtained by the simulation of the process cannot
reliably be used for estimating $\Delta F$. This is a typical
example of how the lack of knowledge of the tails of the work
distributions in micro-manipulations experiments hinders the
possibility of using (\ref{jarzder}) to evaluate free energy
differences.

\begin{figure}[ht]
\psfrag{m}[ct][ct][1.]{\large $m_\C^*$}
\psfrag{t}[ct][ct][1.]{\large $t$}
\psfrag{w}[ct][ct][1.]{\large $w$}
\psfrag{l}[ct][ct][1.]{\large $\lambda^*$}
\psfrag{l1}[ct][ct][1.]{$\lambda$}
\begin{center}
\includegraphics[width=7cm]{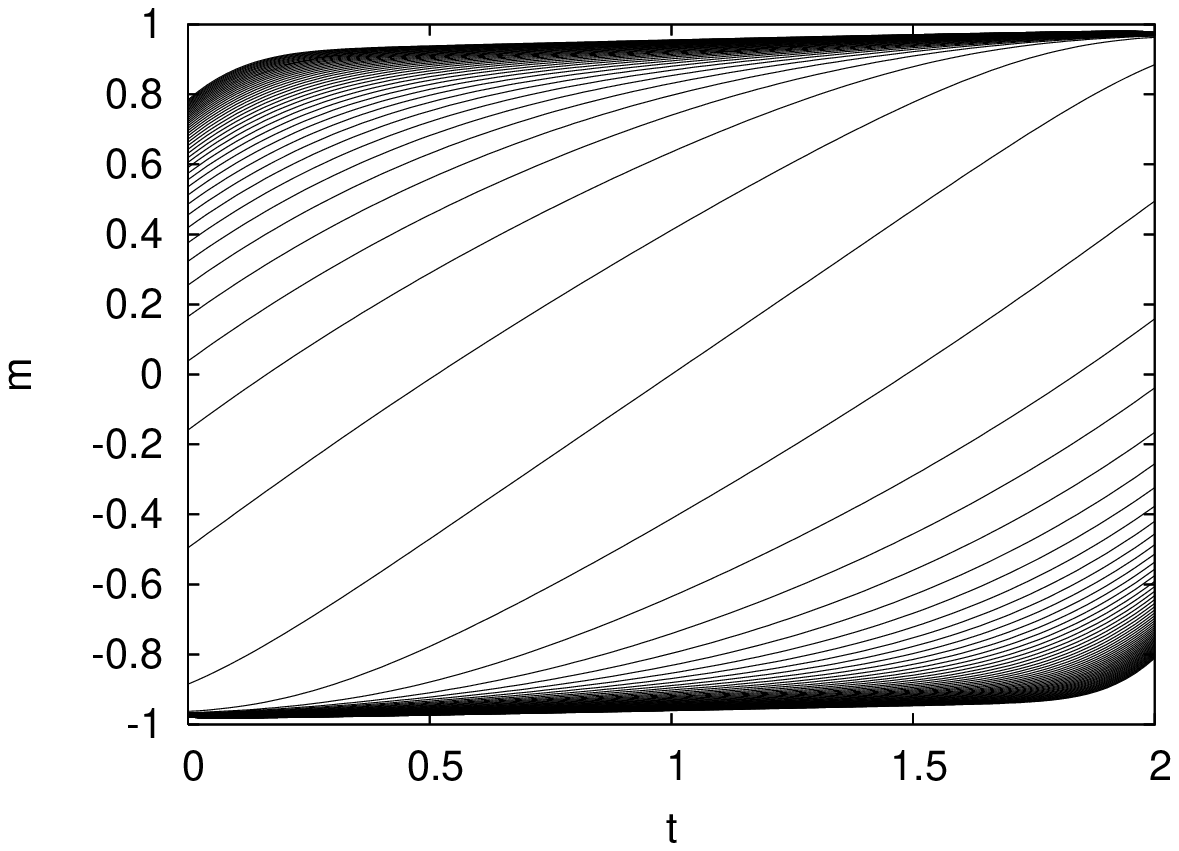}
\includegraphics[width=7cm]{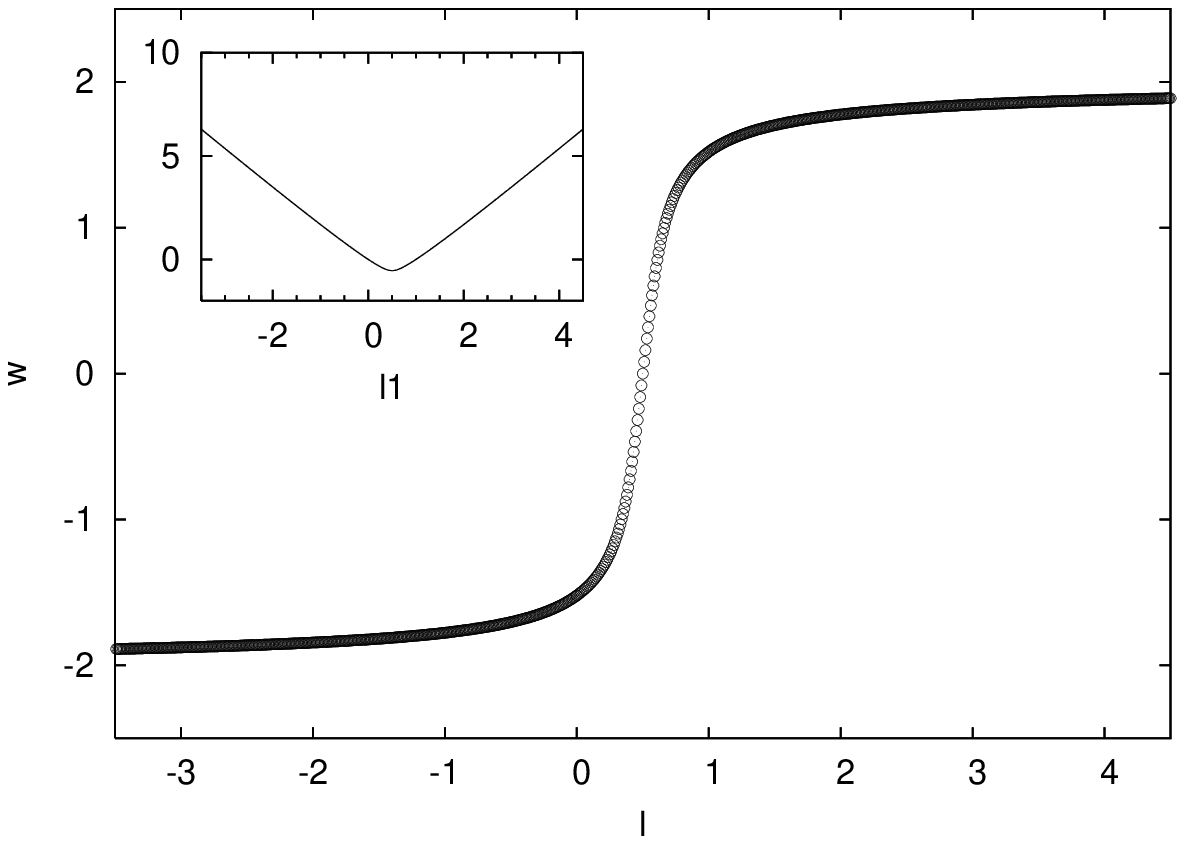}
\end{center}
    \caption{Left: plot of $m_\C^*$ as a function of $t$ for
    different values of $\lambda$,  with $J=1.1$, $h_0=-h_1=-1$,
    and $t_\mathrm{f}=2$.
    The values of $\lambda$ vary between $\lambda=-5$ (bottom curve)
and $\lambda=5$ (top curve),
    with a step  $\Delta \lambda=0.2$. Right:
plot of $w$ as a function of  $\lambda^*$ ,
    as defined by (\ref{eqw:eq}). Inset: plot of $g$ as a function of $\lambda$ as given by (\ref{defgibbs}).}
\label{mt_j1.1}
\end{figure}

\begin{figure}[ht]
\center
\psfrag{m}[ct][ct][1.]{\large $m_\C^*$}
\psfrag{t}[ct][ct][1.]{\large $t$}
\psfrag{w}[ct][ct][1.]{\large $w$}
\psfrag{l}[ct][ct][1.]{\large $\lambda^*$}
\psfrag{g}[cl][cl][1.]{$g$}
\psfrag{l1}[ct][ct][1.]{$\lambda$}
\includegraphics[width=7cm]{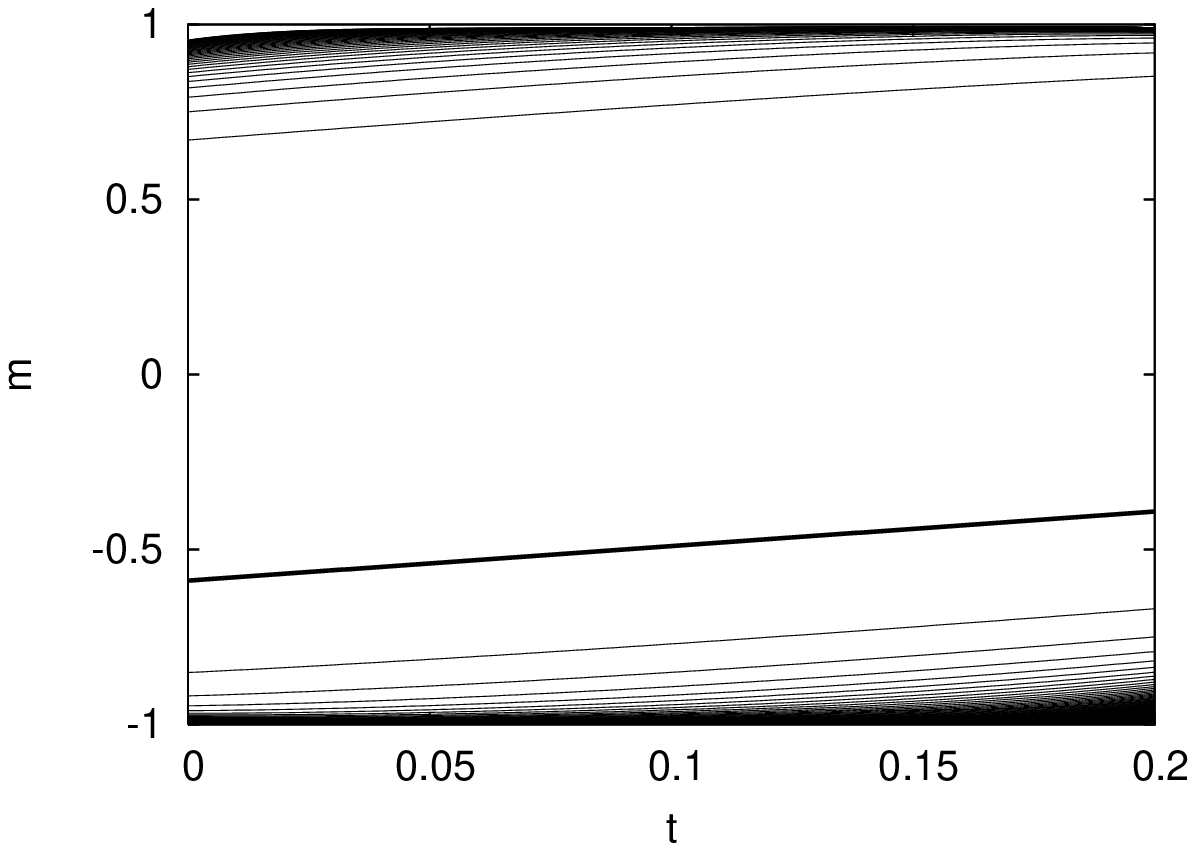}
\includegraphics[width=7cm]{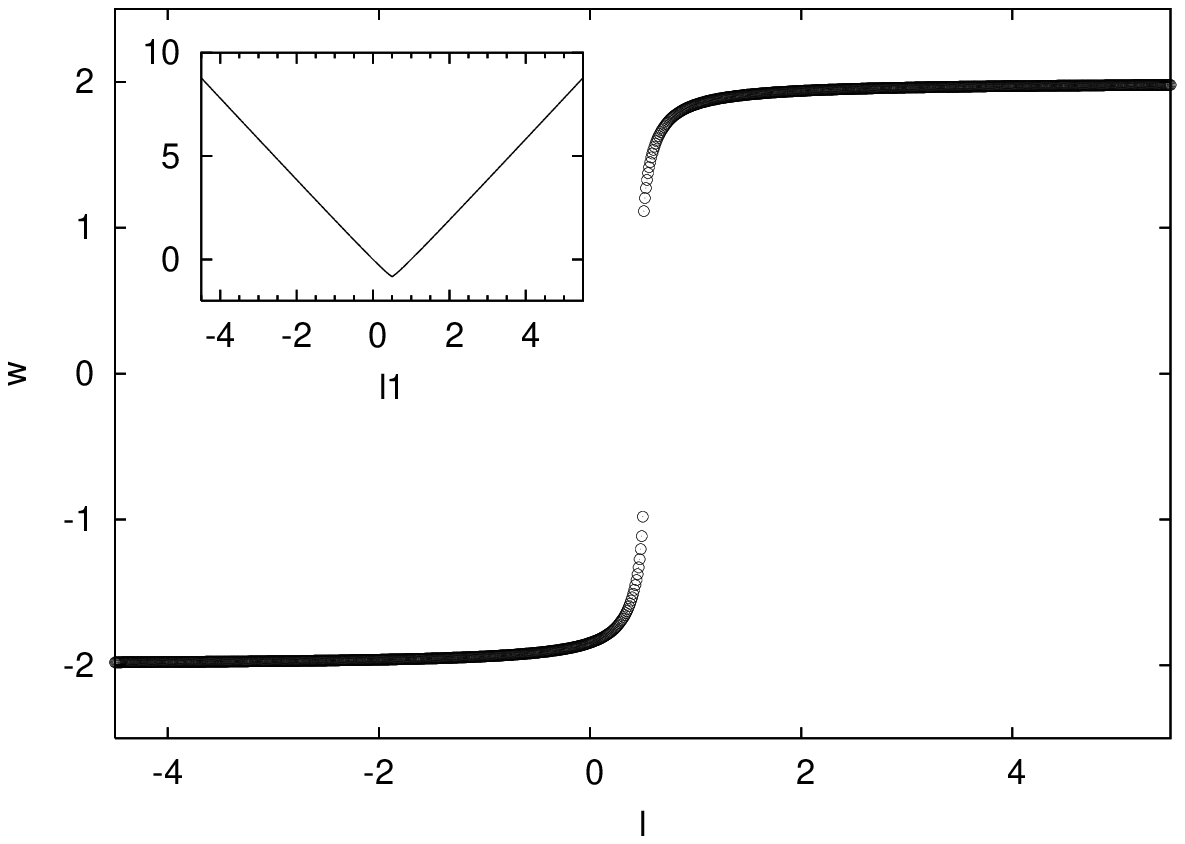}
    \caption{Left: Plot of $m_\C^*$ as a function of $t$ for different
        values of $\lambda$,  with $J=1.1$, $h_0=-h_1=-1$,
        and $t_\mathrm{f}=0.2$.
        The values of $\lambda$ vary between
        $\lambda=-5$ (bottom curve) and $\lambda=5$ (top curve),
        with a step  $\Delta \lambda=0.2$.
        Thick line: $\lambda=0.5$. Right:  plot of $w$ as a function
    of $\lambda^*$, as defined by (\ref{eqw:eq}). Inset: plot of $g$ as a function of $\lambda$ as defined by (\ref{defgibbs}).}
\label{mt1_j1.1}
\end{figure}

\begin{figure}[ht]
\center
\psfrag{J0}[ct][ct][1.]{$J$}
\psfrag{r}[ct][ct][1.]{$r$}
\includegraphics[width=7cm]{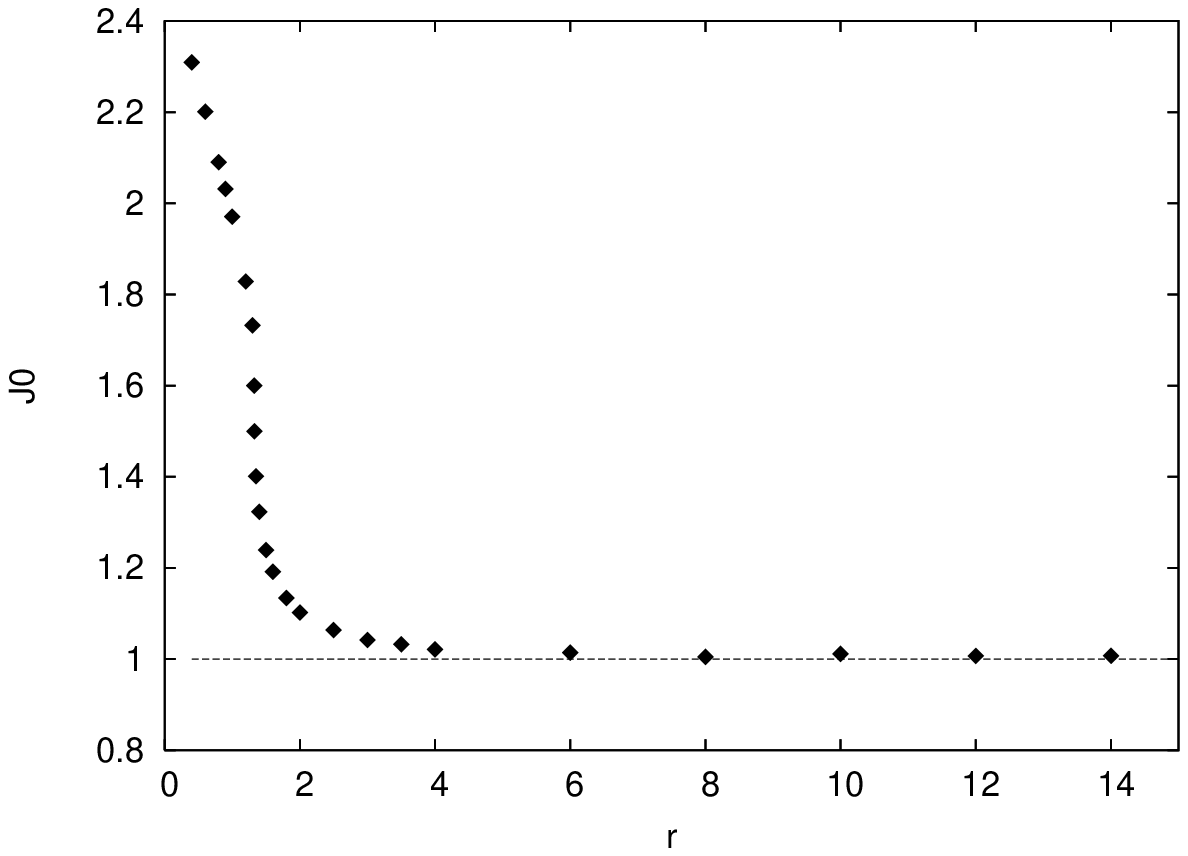}
\psfrag{p}[ct][ct][1.]{\large $\phi$}
\psfrag{w}[ct][ct][1.]{\large $w$}
\includegraphics[width=7cm]{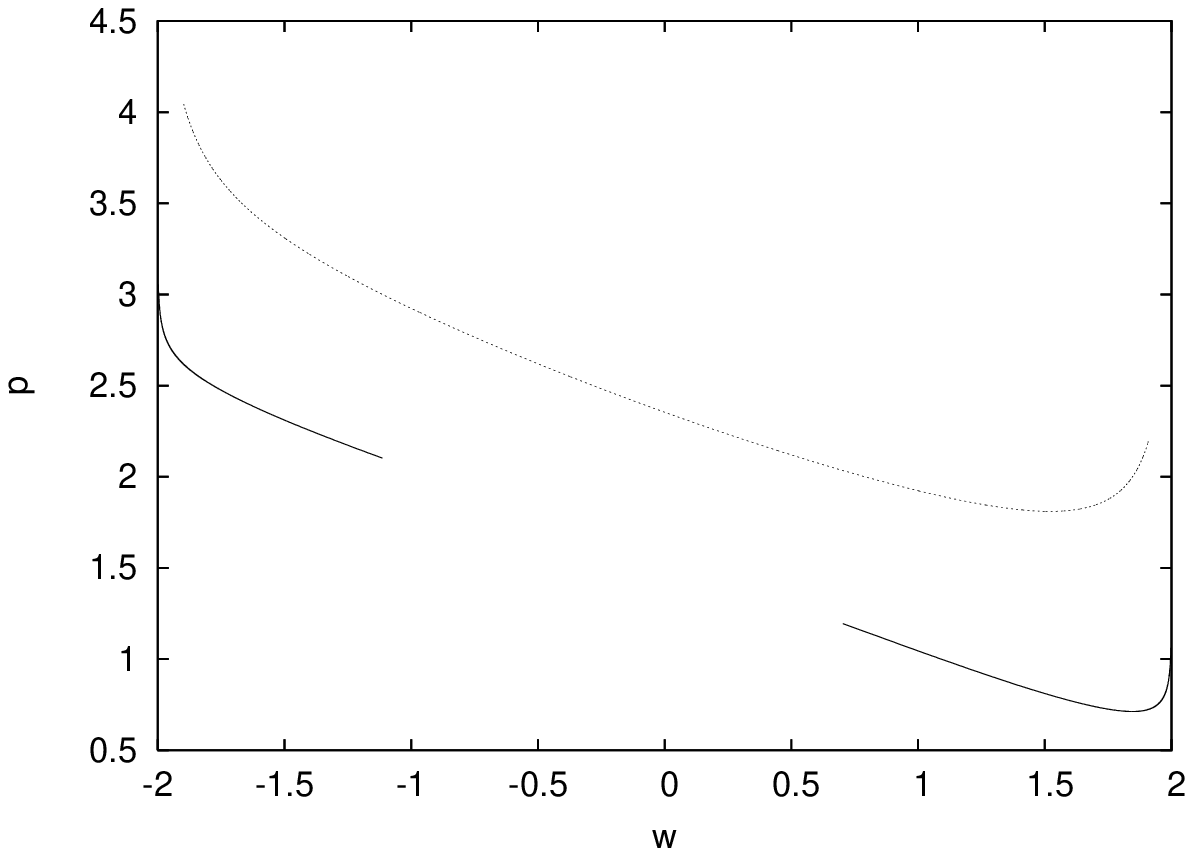}
\caption{Left: Ising model phase diagram, in the $r=(h_1-h_0)/\tf,\, J$ plane. In the upper right part of the diagram
the system exhibits  path phase separation, while in the lower left part no phase separation is
found. The points correspond to the value $J^*(r)$
such that for $J>J^*( r)$ the function  $w(\lambda^*)$ exibits a discontinuity at $\lambda=1/2$. Right: Path Helmholtz free energy  for the system described by
the differential operator (\ref{FP}), manipulated according to the
protocol (\ref{manh:eq}). Full line: $J=1.1$, $h_0=-h_1=-1$, and $t_\mathrm{f}=0.2$.  Dotted line: $J=1.1$, $h_0=-h_1=-1$, and $t_\mathrm{f}=2$. The curves are arbitrarily shifted for clarity.}
\label{fase}
\end{figure}

We now consider the classical paths $m_\C^*(t,\lambda)$, solutions of equations~(\ref{eq2_lang}), and thus contributing to the probability distribution $P(w)$ via~(\ref{eqw:eq}).
In the left panel of
figure~\ref{mt_j1.1}, we plot $m_\C^*$ as a function of $t$ for
different values of $\lambda$, obtained by numerical solution of
equations~(\ref{eq2_lang}), for
$t_\mathrm{f}=2$: we observe that the trajectory
 $m_\C^*(t,\lambda)$ varies continuosly as $\lambda$ is varied. As a consequence,  since the work done on the system along each trajectory  $m_\C^*(t,\lambda)$ reads $w=-\int_0^\tf \D t' \dot h(t') m_\C^*(t',\lambda)$, $w$ turns out to be a continuous function of $\lambda$
(figure~\ref{mt_j1.1}, right panel).
Furthermore,  since $w$ and the saddle point value $\lambda^*$ are related by (\ref{eqw:eq}),
 the function $g(\lambda)$ as given by (\ref{defgibbs})
is a differentiable function with respect to $\lambda$, as shown in the inset of  figure~\ref{mt_j1.1}, right panel.
We now consider a faster protocol, $t_\mathrm{f}=0.2$: the results are plotted in figure \ref{mt1_j1.1}.
One can clearly see that the classical paths $m_\C^*(t,\lambda)$ exhibit
a discontinuity for $\lambda=0.5$, jumping from negative to
positive values. Accordingly, $w(\lambda^*)$ exhibits a
discontinuity at $\lambda^*=0.5$, as shown in the right panel of
figure~\ref{mt1_j1.1}.
This is reflected in the
appearence of a cusp in the function $g(\lambda)$, at $\lambda=1/2$, see the inset of figure~\ref{mt1_j1.1}.
We find that for any value of $r=(h_1-h_0)/\tf$ it is always possible to find a value of $J=J^*(r)$
such that for $J>J^*( r)$ the function  $w(\lambda^*)$ exibits a discontinuity at $\lambda=1/2$,
and thus the trajectories $m^*_\C(t,\lambda)$ exhibit a path phase separation.
In figure~\ref{fase}, left panel, we show the system phase diagram: in the upper right part of the diagram
the system exhibits  path phase separation, while in the lower left part the trajectories $m^*_\C(t,\lambda)$ vary continuously as $\lambda$ is varied.

Let us now define the function
\begin{equation}\label{pathhelm:eq}
    \phi(w)=-\lim_{N\to\infty}\frac{1}{N}\log P(Nw,\tf).
\label{defhelm}
\end{equation}
The functions $g(\lambda)$ and $\phi(w)$
are related by a Legendre transformation:
\begin{equation}\label{legendre:eq}
    \phi(w)=\inf_\lambda \left(
    -g(\lambda)+\lambda w\right)=-g(\lambda^*(w))+\lambda^*(w) \,w,
\end{equation}
where $\lambda^*(w)$ is the solution of (\ref{eqw:eq}). Thus
$\lambda$ and $w$ act like thermodynamically conjugate
variables.
These functions
 can be thus interpreted in terms of
path thermodynamics: $g(\lambda)$ can be viewed as a path Gibbs free
energy, while $\phi(w)$ is the corresponding Helmholtz free energy.
This analogy was first pointed out in~\cite{Ritort2state04,Imparato05}.

How do we interpret the singularity in $g(\lambda)$?
In strict analogy with thermodynamics,
the discontinuity of $g'(\lambda)$ with respect to its independent variable
$\lambda$ corresponds to the linear behavior of its Legendre
transform $\phi(w)$ between $(w_+,\phi(w_+))$ and $(w_-,\phi(w_-))$,
 where $w_\pm$ are the
values of $w$ either side of the discontinuity. Thus the ``path
phase coexistence'' appears as an exponential dependence of
$P(Nw,t_\F)=\exp(-N\phi(w))$ on $W$ in a certain interval.
The presence of such exponential tails was conjectured
in~\cite{Ritort2state04} for a system of independent spins,
but we were only able to exhibit them in an interacting system
like the present one.


\section{The distribution of heat flow in a Markov process}
In this section we discuss the equation governing the time evolution of the probability distribution function of the entropy which flows into the
enviroment surrounding a stochastic system which evolves across its phase space.
We assume that the system at issue has a discrete phase space and its time evolution  is a  stochastic markovian process described by (\ref{evp}).
For simplicity,
we consider a stochastic dynamics with a discrete small time scale
$\tau$, such that the jumps between states take place at discrete
times $t_k=k \tau$.
We consider a generic path
$\omega$ defined by $\omega(t)=i_k$ iff $t_k\le t < t_{k+1}$, with
$k=0,1,\ldots,M$, with $t_{M+1}={\tf}$, and define the
time-reversed path $\widetilde\omega$ by $\widetilde\omega(t)=i_k$
for $\tilde t_{k+1}\le t <\tilde t_{k}$, where $\tilde
t=t_0+{\tf}-t$.
Let us define the quantity $Q(\omega)$ by
\begin{equation}
 Q(\omega)=-\ln
    \left[\frac{\PP(\omega)}{\widetilde\PP(\widetilde\omega)}\right]
    =-\sum_{k=1}^M \ln \left[\frac{K_{i_{k}i_{k-1}}(t_k)}{K_{i_{k-1}
    i_{k}}(t_k)}\right],
\label{entrflow_def}
\end{equation}
where $\PP(\omega)$ is the probability of the forward path $\omega$
(conditioned by its initial state $i_0$) and
$\widetilde \PP(\widetilde \omega)$ is the probability of the time-reversed path
$\widetilde \omega$, conditioned by \textit{its} initial state
$i_M\equiv i_{\F}$ and subject to the time-reversed protocol
$\widetilde K_{ij}(t)=K_{ij}(\tilde t)$~\cite{LebSpohn99,Crooks99,Crooks00,Seifert05}.
We have assumed that, if $K_{ij}(t)>0$ at any time $t$, one also has
$K_{ji}(t)>0$.

It is worth noting that, if the detailed balance conditions holds for the
transition rates $K_{ij}(t)$, and the energy $H_i(t)$ is associated
to the state $i$ of the system, we have
$K_{ji}(t)/K_{ij}(t)=\exp\left\{
\left[H_i(t)-H_j(t)\right]/T\right\}$, and thus
$T\ln\left[K_{ji}(t)/K_{ij}(t)\right]$ represents the heat exchanged
with the reservoir in the jump from state $j$ to state $i$. (In this section
we set $\kb=1$.)
Thus the
quantity $Q(\omega)$, defined by~(\ref{entrflow_def}), is the entropy
which flows into the reservoir as the system evolves along the path
$\omega$~\cite{LebSpohn99,Gaspard04,Gaspard05,Imparato06c}.
Let us define $\Delta s_{ij}$ as the entropy which flows into the reservoir as
a result of the jump of the system from state $j$ to state $i$ $\Delta s_{ij} = \log\left[K_{ji}(t)/K_{ij}(t)\right]$.
The differential equation governing the time
evolution of the joint probability distribution function $\Phi_i(Q,t)$ reads \cite{Imparato06c}
\begin{equation}
    \frac{\partial \Phi_i(Q,t)}{\partial t}=\sum_{j\,(\neq i)}
    \left\{ K_{ij} \left[\sum_{n=0}^\infty
    \frac{\left(-\Delta s_{ij}\right)^n}{n!}
    \frac{\partial^n\Phi_j(Q,t)}{\partial Q^n} \right]
    -K_{ji}\Phi_i(Q,t)\right\}.
\label{eq_phi}
\end{equation}
By introducing, for each $i$, the generating function
$\Psi_i(\lambda,t)=\int \D Q\; \exp(\lambda Q)\, \Phi_i(Q,t)$,
and taking into account the expression of $\Delta s_{ij}$, we obtain
the master equation
\begin{equation}
    \frac{\partial\Psi_i(\lambda,t)}{\partial t}=\sum_{j\,(\ne i)}
    \left[K_{ij} \left(\frac{K_{ji}}{K_{ij}}\right)^\lambda
    \Psi_j(\lambda,t) -K_{ji}\Psi_i(\lambda,t)\right]
    =\sum_j H_{ij}(\lambda)\,\Psi_j(\lambda,t) . \label{master}
\end{equation}
which was first derived by Lebowitz and Spohn in~\cite{LebSpohn99}.
In the case of time-independent transition rates $K_{ij}$,
or of
transition rates which depend periodically on the time, it can be
useful, in order to evaluate the distribution function $\Phi(Q,t)=\sum_i\Phi(Q,t)$,
to introduce the large-deviation function. In the long-time limit,
the generating function $\Psi(\lambda,t)=\sum_i\Psi_i(\lambda,t)$ is dominated by the
maximum eigenvalue $g(\lambda)$ of the matrix
$\mathsf{H}(\lambda)=\left(H_{ij}(\lambda)\right)$, which appears in
the master equation (\ref{master}). Therefore, we have, for long
times $t$,
\begin{equation}
\Psi(\lambda,t)\propto\exp\left[t \,g(\lambda)\right]. \label{gf}
\end{equation}
By using the last equation and using the definition of the generating function,  one obtains the
probability distribution of the entropy flow in the long time limit:
\begin{equation}
\Phi(Q,t)=\int \frac{\D \lambda}{2\pi\I}\; \E^{-\lambda Q}
\Psi(\lambda, t)
    \propto \E^{t\,g(\lambda^*)-\lambda^* Q}, \label{large_phi}
\end{equation}
where $\lambda^*$ is the saddle point value implicitly defined by
$\left.\partial g/\partial \lambda\right|_{\lambda^*}=Q/t$. If we
introduce the entropy flow per unit time $q=Q/t$, we obtain the
large-deviation function
\begin{equation}
    f(q)\equiv g(\lambda^*)-\lambda^* q
    =\lim_{t\rightarrow\infty} \frac 1 t \log \Phi(t q,t).
\label{fq_def}
\end{equation}
Note that the functions $g(\lambda)$ and $f(q)$ are Legendre
transform of each other, and can be then interpreted in terms of
path thermodynamics: $g(\lambda)$ can be viewed as a path Gibbs free
energy, while $f(q)$ is the corresponding Helmholtz free energy.
The connection between the generating function $\Psi(\lambda)$ and the thermodynamic formalism for dynamical systems \cite{Ruelle:book} has been investigated in Ref. \cite{lecomte-2005}.

If the system is characterized by a small number of states, one can
explicitly solve the equations~(\ref{master}), and thus obtain the
total generating function $\Psi(\lambda,t)\equiv\sum_i
\Psi_i(\lambda,t)=\average{\E^{\lambda Q}}.$
While on the one hand this direct approach becomes rapidly impracticable, as the
system phase space size increases,
on the other hand (\ref{master}) suggests a practical
computational scheme to evaluate the  generating function $\Psi(\lambda,t)$.
 Since
$\Psi(\lambda,t)=\int \mathcal{D}\omega_t\; \PP(\omega_t)\,
\E^{\lambda Q(\omega_t)}$, we have
\begin{equation}
\frac{\partial\Psi(\lambda,t)}{\partial \lambda}
=\average{Q}_\lambda \Psi(\lambda,t), \label{psidel}
\end{equation}
where $\average{\dots}_\lambda$ is the average in the weighted
ensemble $\PP(\omega_t) \exp\left[\lambda
Q(\omega_t)\right]/\Psi(\lambda,t)$, where
$\Psi(\lambda,t)=Z_\lambda$ is the ``partition function'' of this
ensemble, which will be called the ``$\lambda$-ensemble'' in the
following. The solution of~(\ref{psidel}) thus reads
\begin{equation}
\Psi(\lambda,t)=\exp\left[\int_0^\lambda\D \lambda'
\average{Q}_{\lambda'}\right]. \label{psi_sun}
\end{equation}

Following \cite{Imparato06c}, in the present paper
we consider a procedure to evaluate $\Psi(\lambda,t)$ which generates trajectories in a
suitable entropy-flow weighted ensemble. The direct simulation of
trajectories in the $\lambda$-ensemble is hindered by the fact that
one should already know the exact expression of the ensemble
partition function, i.e., the function $\Psi(\lambda,t)$, which is
the unknown quantity at issue.

To avoid the problem of the direct evaluation of $\Psi(\lambda,t)$,
following~\cite{Oberhofer05}, we introduce a generic functional of the paths
$\Pi(\omega)$, and write
\begin{equation}
\average{Q}_\lambda=\frac{\int \mathcal{D}\omega_t\;
(Q(\omega_t)/\Pi(\omega_t))\Pi(\omega_t) \PP(\omega_t)\E^{\lambda
Q(\omega_t)}} {\int \mathcal{D}\omega_t\;
(1/\Pi(\omega_t))\Pi(\omega_t) \PP(\omega_t)\E^{\lambda Q(\omega_t)}
}
=\frac{\average{Q/\Pi}_{\lambda,\Pi}}{\average{1/\Pi}_{\lambda,\Pi}},
\label{qlp}
\end{equation}
where $\average{\dots}_{\lambda,\Pi}$ indicates the average in the
new $\PP(\omega_t) \Pi(\omega_t)\exp\left[\lambda
Q(\omega_t)\right]$ ensemble, which will be indicated as the $(\lambda,\Pi)$-ensemble in the
following.

We choose the functional of the path $\Pi(\omega)$ as discussed in \cite{Imparato06c}.
The probability of a given path $\omega$ reads
\begin{equation}
\PP(\omega)=\Omega_{i_N,i_{N-1}}(t_{N-1}) \Omega_{i_{N-1},i_{N-2}}(t_{N-2})\dots \Omega_{i_1,i_{0}}(t_0)p^0_{i_0},
\end{equation}
where the transition probabilities $\Omega_{i,j}(t)$ are defined as
$\Omega_{i,j}=\tau K_{i_j}(t)$, and $\Omega_{i,i}=1-\sum_{j (\ne i)}\Omega_{ji}(t)$. We
now define the new transition probabilities $\widetilde \Omega_{i,j}=\tau
K_{ij} \left(K_{ji}/K_{ij}\right)^\lambda$, and $\widetilde
\Omega_{i,i}=1-\sum_{j (\ne i)}\widetilde \Omega_{ji}$, and choose the
functional $\Pi(\omega)$, such that \cite{Imparato06c}
\begin{equation}
\Pi(\omega)=\prod_{k=1}^{M}\Pi_{i_k,i_{k-1}}(t_k),\quad \mbox{ with }\quad
\Pi_{ij}(t)= \cases{
    1,
    & if $i\ne j$ ;\cr
   \widetilde \Omega_{jj}(t)/ \Omega_{jj}(t),  & if
   $i=j$.}
\label{pij}
\end{equation}
Recalling the definition of $Q(\omega)$, (\ref{entrflow_def}), we
obtain that the probability in the $(\lambda,\Pi)$-ensemble is given
by
\begin{equation}
\PP(\omega)\Pi(\omega) \exp\left[\lambda Q(\omega)\right]=
\widetilde \Omega_{i_N,i_{N-1}}\widetilde  \Omega_{i_N,i_{N-1}}\dots
\widetilde \Omega_{i_1,i_{0}}p^0_{i_0}, \label{pp_pi_q}
\end{equation}
and thus $\average{Q}_\lambda$ can be evaluated by using~(\ref{qlp}),
for the particular choice of $\Pi$, as given by (\ref{pij}). Note
that (\ref{pp_pi_q}) implies that the one can generate a trajectory
in the $(\lambda,\Pi)$-ensemble by simply simulating the process with
the $\widetilde \Omega_{i,j}(t)$ transition probabilities.

In \cite{Imparato06c}, the
the feasibility of the method was illustred by applying it to  a nonequilibrium system characterized by
a large phase space, and evolving according to
a stochastic dynamics, namely the simple
asymmetric exclusion process (ASEP)~\cite{SchutzDomanyASEP93}.
Such a system  consists in a
one-dimensional lattice gas on a lattice of $L$ sites. Each site of
the model is either empty or occupied by at most one particle. Each
particle can jump into an empty nearest neighbor site with
transition rates per time unit $K_+$ (rightward) and $K_-$
(leftward). The system is kept in an out-of-equilibrium steady state
since its first and last site are in contact with two particle
reservoirs, at densities $\rho_A$ and $\rho_B$ respectively. By
taking $\rho_A>\rho_B$ and $K_+>K_-$, one observes a net particle
current from the left to the right reservoir.
In \cite{Imparato06c} we considered an ASEP model with constant parameters $\rho_A,\, \rho_B,\, K_+,\, K_-$. Here we consider a system with time-dependent parameters.
We take $K_+=1$, $K_-=K_-^0  (1+\sin(2\pi t))+\epsilon$, with $K_-^0=0.75,\, \epsilon=10^{-3}$,\, $\alpha=1,\, \gamma=0.27,\,  \beta=1,\, \delta=0.27$, where
$\alpha$-$\gamma$ are the rates of jump from-into the  left $A$ reservoir, respectively, and
$\beta$-$\delta$ are the rates of jump into-from the right $B$ reservoir, respectively.
With this choice of parameters, at $t=2 k \pi$, we have
$\rho_A=0.75$, $\rho_B=0.25$~\cite{ASEP04}.
The duration of a  single trajectory is taken to be $\tf=10 \cdot 2 \pi$.
\begin{figure}[ht]
\center \psfrag{Gamma}[ct][ct][1.]{$g$ }
\psfrag{l}[ct][ct][1.]{$\lambda$ }
\includegraphics[width=7cm]{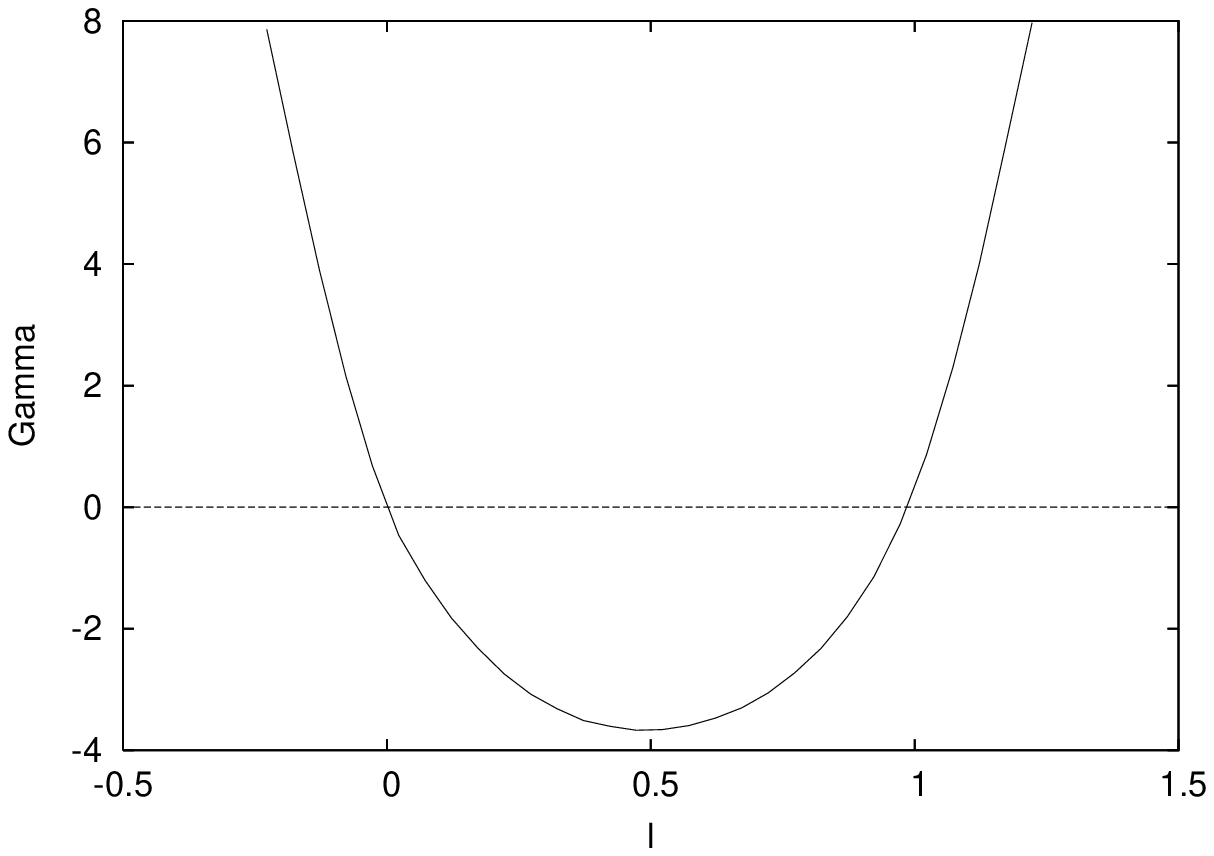}
\psfrag{P}[ct][ct][1.]{$\Phi$}
\psfrag{f}[lb][lb][1.]{$\Phi$}
\psfrag{q}[ct][ct][1.]{$q$}
\includegraphics[width=7cm]{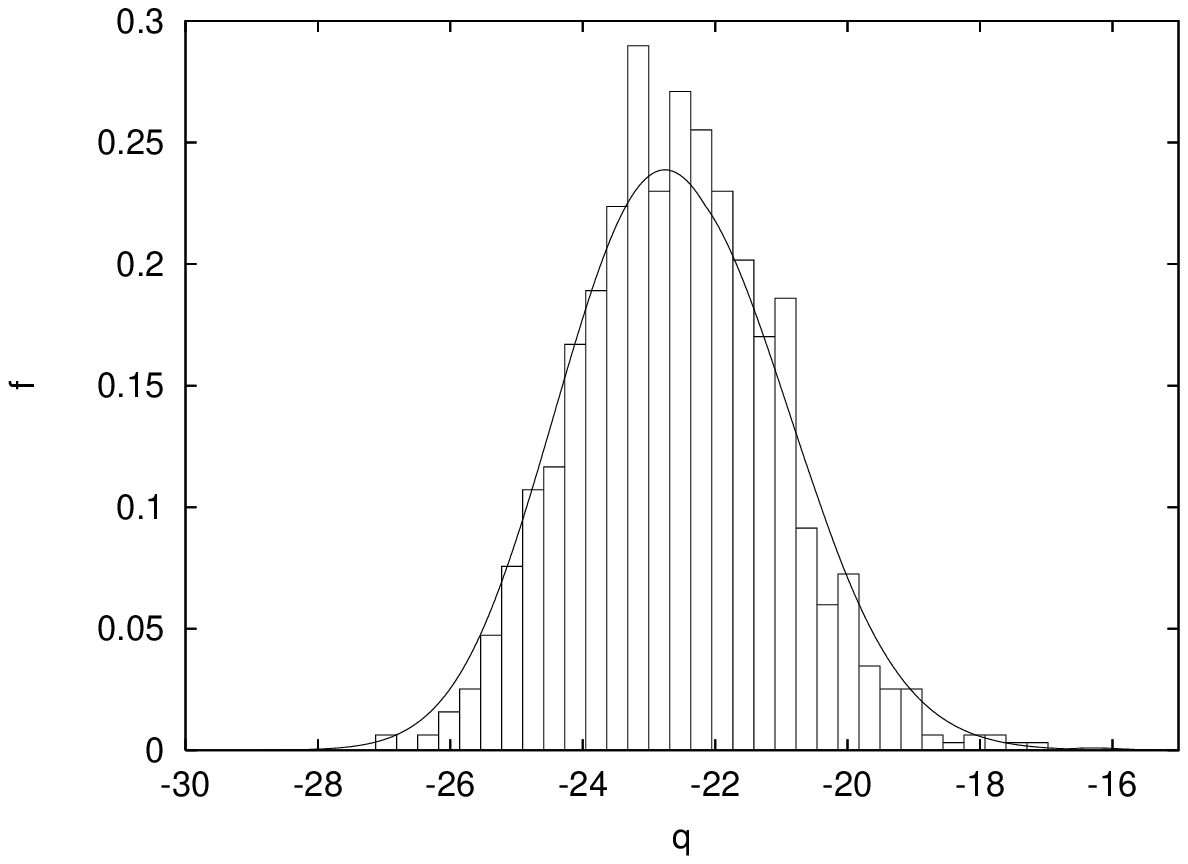}
\caption{Left: Plot of $g(\lambda)$ as obtained by combining~(\ref{gf}) and
(\ref{psi_sun}), for the ASEP model. The function $g(\lambda)$
vanishes for $\lambda=0,1$ which corresponds to the normalization
condition and to Seifert's fluctuation relation \cite{Seifert05}
respectively. Right: Histogram of the entropy flow per time unit $q$,
corresponding to 1000 unbiased trajectories. Full line: probability
distribution function $\Phi(t_\F q,{t_\F})$, obtained with the trajectory
sampling algorithm.} \label{fig1}
\end{figure}
A direct evaluation of the function $g(\lambda)$ by solving the
$2^{100}$ equations (\ref{master}) is out of question. We thus
apply our trajectory simulation approach to the ASEP model. We
consider trajectories  with elementary
time step $\tau=0.01$: at each time the transition probability
between two states is given by the transition matrix $\widetilde
K_{ij}(t)$. For each value of $\lambda$ we generate
$\mathcal{N}=10000$ sample trajectories and calculate the entropy
flow $Q$, as defined by~(\ref{entrflow_def}), for each trajectory.
Then, for the given value of $\lambda$, by averaging over the
$\mathcal{N}$ trajectories, we compute the quantity
$\average{Q}_\lambda$ using~(\ref{qlp}). Finally, by combining
(\ref{gf}) and~(\ref{psi_sun}), we obtain the function $g(\lambda)$
which governs the long time behavior of $\Psi(\lambda,t)$. This
function is plotted in figure~\ref{fig1}, left panel. It can be seen that it vanishes
at $\lambda=0,1$ and is symmetric with respect to $\lambda=1/2$. The
fact that $g(0)=0$ corresponds trivially to the normalization
condition over all the possible trajectories. On the other hand, the
fact that the function $g$ vanishes at $\lambda=1$, is a non trivial
result, and corresponds to Seifert's fluctuation theorem
\cite{Seifert05}. The symmetry around $\lambda=1/2$ corresponds to
the Gallavotti-Cohen fluctuation relation. We are now able to
calculate the large-deviation function $f(q)$ defined
in~(\ref{fq_def}). We check as follows that
the quantity $f(q)$ actually gives the entropy distribution function
$\Phi(q,t)\propto \exp\left[t\, f(q)\right]$ for the present model.
We simulate the unbiased diffusion process by using the
transition matrix $K_{ij}(t)$, and measure the entropy flow along 1000
trajectories. We then plot the histogram of the measured entropy
flow per time unit, together with the function $ \exp\left[t\,
f(q)\right]$, see figure~\ref{fig1}, right panel. The agreement between the histogram
and the predicted entropy distribution $\Phi(q,t)$ is excellent.

\section{Discussion}
We have seen that the equations governing the evolution of the work
and heat flow distributions in out-of-equilibrium systems can be
harnassed to yield interesting information. We have analyzed the
generating function of the work distribution in ``large''
manipulated systems and shown that it may exhibit a ``path phase
transition'', which corresponds to the presence of an exponential
behavior in some work interval. On the other hand, we have shown
that it is possible to evaluate the large deviation function for the
entropy flow in a stochastic process via a biased simulation
technique, and applied the method to exhibit the Gallavotti-Cohen
symmetry in an out-of-equilibrium system with periodically varying
parameters. A similar result has been recently obtained by Ge and
collaborators~\cite{Ge06}.


\section*{Acknowledgements}
We are grateful to Pierre Gaspard and Christian van~den~Broeck for
having given us the opportunity of taking part in this exciting meeting.
We also thank C. Jarzynski and U. Seifert for their interest in our work.

\bibliographystyle{elsart-num}

\bibliography{fluctheor}

\end{document}